\documentclass[twocolumn,showpacs,preprintnumbers,amsmath,amssymb]{revtex4}

\usepackage{CJK}
\usepackage{graphicx}
\usepackage{dcolumn}
\usepackage{bm}

\def\btbl{\begin{tabular}} \def\etbl{\end{tabular}}
\def\bcc{\begin{center}} \def\ecc{\end{center}}
\def\beq{\begin{equation}} \def\eeq{\end{equation}}
\def\btbl{\begin{tabular}} \def\etbl{\end{tabular}}
\def\NUFM{{\footnotesize NUFM}} 
 
\def\CERN{{\footnotesize CERN}} \def\SPS{{\footnotesize SPS}}
\def\LHC{{\footnotesize LHC}} \def\QGP{{\footnotesize QGP}}
\def\E941{{\footnotesize E941}} \def\E864{{\footnotesize E864}}
\def\NA49{{\footnotesize NA49}} \def\NA35{{\footnotesize NA35}}
\def\RHIC{{\footnotesize RHIC}} 
  \def\Boltzmann{{\footnotesize Boltzmann}}
\def\GeV{{\footnotesize GeV}} 
\def\BNL{{\footnotesize BNL}} \def\AGS{{\footnotesize AGS}}
\def\RHIC{{\footnotesize RHIC}} \def\LHC{{\footnotesize LHC}}

 \def\MeV{{\footnotesize MeV}}

\def\AGS{{\footnotesize AGS}}

\begin{document}
\title{Baryon productions and collective flow of relativistic
heavy-ion collisions in the AGS, SPS, RHIC and LHC energy regions($\sqrt{s_{NN}}\leq5$GeV to $5.5$ TeV) }

\author{Shengqin ~Feng$^{1,2,3}$}
\author{Yang Zhong$^1$}

\affiliation{$^1$College of Science, China Three Gorges University,
Yichang 443002, China} \affiliation{$^2$Key Laboratory of
Quark and Lepton Physics (Huazhong Normal Univer.), Ministry of
Education£¬Wuhan 430079£¬China} \affiliation{$^3$School of Physics
and Technology, Wuhan University, Wuhan 430072, China}

\begin{abstract}
The features of net baryon productions and collective flow in relativistic
heavy-ion collisions at energies reached at the CERN Large Hadron
Collider (LHC), BNL Relativistic Heavy Ion Collider (RHIC) ,
CERN Super Proton Synchrotron (SPS) and BNL Alternating Gradient Synchrotron (AGS)
with the model of Non-Uniform Flow Model(NUFM) are systematically studied in this paper.  Especially we
predict the feature of net baryon productions and collective flow at LHC $\sqrt{s_{NN}}$=5500 GeV
basing on the detailed study of that at RHIC  $\sqrt{s_{NN}}$=62.4 and 200GeV.
The dependencies of the features of baryon
stopping and collective flow on the collision energies and centralities are
investigated.\\

\vskip0.2cm \noindent Keywowds: ~~Non-Uniform Flow
Model,
~~LHC, ~Collective flow, ~net proton distributions
\end{abstract}

\pacs{25.75.-q, 25.75.Ld, 25.75.Dw} \maketitle

\section{Introduction}
\label{intro}

Over the last two decades, relativistic heavy ion collisions~\cite{Mueller,Jacobs} have been studied
experimentally at increasingly higher center-of-mass energies at the Brookhaven
Alternating Gradient Synchrotron AGS ($\sqrt{s_{NN}}<5$  GeV), the ~\CERN~ Super Proton
Synchrotron ~\SPS~ ($\sqrt{s_{NN}}\leq 20$ GeV) and the Brookhaven Relativistic Heavy Ion Collider
RHIC ($\sqrt{s_{NN}}\leq 200$ GeV). As discussed in this article, the data collected in these
experiments display remarkable generic trends as a function of system size
and kinematic variables. The Large Hadron Collider LHC at CERN will study heavy ion collisions at a
center-of-mass energy $\sqrt{s_{NN}}$ = 5.5 TeV, which is a factor 27 higher than the
maximal collision energy at \RHIC. This
is an even larger increase in center of mass energy than the factor
10 in going from the \CERN~ \SPS~ to \BNL~ \RHIC. It leads a significant
extension of the kinematic range in longitudinal rapidity and
transverse momentum. The collectivity of high energy density matter
is one of the important properties to understand high-energy
heavy-ion collisions~\cite{Armesto,Cooper,Landau}. It is also challenging to
understand how collectivity is generated during collisions.

There has been a lot of work in recent years on
thermal and collective flow model
calculations~\cite{Teaney,Hirano,Gyulassy,Kharzeev,Becattini,Rischke,Sinyukov,Bass,
Srivastava,Kolb,Schnedermann,Braun,Bjorken} of heavy ion collisions to \RHIC~ data and extrapolating them to the higher \LHC~ energies.
Here we should mention some kinds of models of thermal and collective flow. The first one is the spherically-expanding source model that may
be expected to approximate the fireball of an isotropic thermal
distribution created in lower-energy collisions.

As the collision energy increases, stronger longitudinal flow is formed which
leads to a cylindrical geometry according to the second kind model~\cite{Schnedermann,Braun}.
It accounted for the anisotropy of longitudinal and transverse direction by adding
the contribution from a set of fireballs with centers located uniformly in the rapidity
region in the longitudinal direction. It can account for the wider rapidity distribution
at \AGS~ and \SPS~ when comparing to the prediction of pure thermal isotropic model.

Bjorken~\cite{Bjorken} postulated that the rapidity distribution of produced particles establishes a
plateau at mid-rapidity which has been formulated
for asymptotically high energies.  It is well
known that collisions at available heavy-ion energy regions of \AGS~,
\SPS~ and \RHIC~ are neither fully stopped nor fully
transparent~\cite{be1,be2,klay,ahle,ba,ap,videbeck,wienold,back},
although a significant degree of transparency is observed. But the
central plateau structure becomes more and more obvious as the
collision energy increases to \SPS~ and \RHIC.

As the collision energy increases to ~\LHC, which is a factor $27$ higher than the
maximal collision energy at \RHIC, the kinematic range in the
longitudinal direction will increase considerably
and the net-baryon density will decrease quickly at mid-rapidity. It seems reasonable to
realize that the plateau proposed by Bjorken ~\cite{Bjorken}
at mid-rapidity at \LHC~ energy region has been established. For the net
baryon distributions, Ref.~\cite{wong} realized  the collision of high-energy heavy ions can
be divided into two different energy regions:
the baryon-free quark gluon plasma(~\QGP) region (or the pure ~\QGP~ region) with $\sqrt{s}>100GeV$ per nucleon, and
the baryon-rich ~\QGP~  region (or the "stopping" region) with $\sqrt{s}\sim5-10GeV$ per nucleon, which
corresponds to about many tens of ~\GeV~ per projectile nucleon in the laboratory system. In the baryon-free ~\QGP~
region, we need to know the nuclear stopping power to determine whether the beam baryons and the target baryons will recede away
from the center of mass without being completely stopped,leaving behind ~\QGP~ with very little baryon content.

The \NUFM~(Non-Uniform Flow Model)~\cite{feng1,feng,feng3,feng4,feng5,feng6}
realized that the fireballs keep some memory on the motion of the
incident nuclei, and therefore the distribution of fireballs,
instead of being uniform in the longitudinal direction, is more
concentrated in the motion direction of the incident nuclei, i.e.
more dense at large absolute value of rapidity. It will not only
lead to anisotropy in longitudinal-transverse directions, but also
render the fireballs (especially for those baryons) distributing
non-uniformly in the longitudinal direction. \NUFM
~\cite{feng1,feng,feng3,feng4,feng5,feng6} may analyze the central dip of baryon
rapidity distribution by assuming that the centers of fireballs are
distributed non-uniformly in the longitudinal phase space.

This paper is organized as follows. In Sec. 2 we give a  brief review the  Non-
Uniform Flow Model in the longitudinal direction. The comparison and analysis of baryon distribution
of \AGS~, \SPS~, \RHIC~ and \LHC~ with the results of the model given in Sec. 3. Section 4
gives a summary and conclusions.

\section{Non-Uniform Flow Model NUFM}

The NUFM model we considered~\cite{feng1,feng,feng3,feng4,feng5,feng6} contains three distinct assumptions:

1.  It is argued that the transparency/stopping of relativistic heavy-ion collisions should be taken into account more carefully.
A more reasonable assumption is that the fireballs keep some memory on the motion of the incident nuclei, and therefore the distribution of fireballs,
instead of being uniform in the longitudinal direction, is more concentrated in the direction of motion of the incident nuclei, i.e. more dense at
large absolute value of rapidity. It will not only lead to anisotropy in longitudinal-transverse directions, but also render the fireballs
(especially for those baryons) distributing non-uniformly in the longitudinal direction.

2. The freeze-out temperatures are assumed to be about the same around 120 ~\MeV~ whether it is at higher \LHC~  or at
lower \AGS~ energy region. Since the temperature at freeze-out exceeds 100 ~\MeV, the ~\Boltzmann~ approximation seems reasonable to study \LHC~ at freeze-out.

3. In order to express the non-uniformity of flow in the longitudinal direction, an ellipticity parameter $e$ is introduced through
a geometrical parametrization. For the central collisions, the nuclear stopping can be studied by the range of rapidity of emission
source in the center-of-mass system.

We have previously used \NUFM~ to study the net proton rapidity among ~\AGS,
\SPS~ and \RHIC~ energy regions~\cite{feng}.  But for the \RHIC~ energy
regions, we  made an earlier error~\cite{feng} to predict the distributions
of net proton distributions since we neglected the effects of the
baryon number conservation. Therefore, it is necessary to reanalyze
the features of net proton rapidity distributions among \AGS~ to
\RHIC~ by taking into account the baryon number conservation. It is
found that when we consider the baryon number conservation, the
features of the distributions at \RHIC~ are completely different from
the results given before~\cite{feng}, especially at large absolute
rapidity region. On the other hands, with the run of forthcoming
\LHC, the predictions of the features of net proton rapidity
distributions at \LHC~ are also important. We will restudy the
features of net proton rapidity distributions among \AGS~ to \RHIC~ by
using \NUFM, and make prediction for the features of forthcoming
\LHC~ in this paper.  In the following, we will firstly make a simple
introduction to the \NUFM.

A parametrization for such a non-uniform distribution can be
obtained by using an ellipse-like picture on emission angle
distribution. In this scenario, the emission angle is:

\begin{equation}  
\theta=\tan^{-1}(\textit{e}\tan\Theta).
\label{eq:eq1} 
\end{equation}

Here, the induced parameter $e(0 \leq e \leq 1)$ represents the
ellipticity of the introduced ellipse which describes the
non-uniform of fireball distribution in the longitudinal
distribution. The detailed discussions of the \NUFM~ were given by
ref.~\cite{feng1}. The rapidity distribution of \NUFM~ is:

\begin{equation}  
\frac{dn_{\textrm{NUFM}}}{dy}=eKm^{2}T\int_{-y_{e0}}^{y_{e0}}\rho(y_{e})dy_{e}(1+2\Gamma+2\Gamma^{2})e^{-1/\Gamma},
\label{eq:eq2} 
\end{equation}

\noindent $y_{e0}$ and $e$ are the important parameters in this paper, $y_{e0}$ is the rapidity limit which confines the rapidity
interval of longitudinal flow and $e$ can describe
the non-uniform in the longitudinal direction of the collective flow. In eq. (2),

\begin{equation}  
\Gamma=T/m\cosh(y-y_{e}),
\label{eq:eq3} 
\end{equation}

\noindent $m$ is the
mass of produced particle, $T$ is the temperature parameter,and $y_{e}$  is the rapidity of collective flow, and

\begin{equation}  
\rho(y_{e})=\sqrt{\frac{1+\sinh^{2}(y_{e})}{1+e^{2}\sinh^{2}(y_{e})}}
\label{eq:eq4} 
\end{equation}

\noindent is the flow distribution function in the longitudinal direction, and $e$ is a parameter which represents the ellipticity
of the introduced ellipse describing the non-uniform of fireball distribution in the longitudinal direction.
It may be figured out
from Eq.2 to Eq.4 that the larger the parameter $e$, the flatter the distribution function $\rho(y_{e})$, the more uniform the
longitudinal flow distribution. When $e\Rightarrow{1}$, the longitudinal flow distribution is completely uniform $\rho(y_{e})\Rightarrow{1}$
and returns to uniform flow. The other important parameter $y_{e0}$ describes the kinematic region and  can determine the width of the distribution.

In order to discuss the dependence of of velocity of collective flow on collision energy in the central mass(CM) system, we
give a calculation of the average velocity
in the longitudinal direction as $<\beta_{L}>=\tanh(y_{e0}/2)$ and $<\beta\gamma>_{L}$, where $\gamma=1/\sqrt{1-<\beta_{L}>^{2}}$ is the
lorentz factor. Therefore
$y_{e0}$ can also determine the average velocity of collective flow in the longitudinal direction.

\section{The net proton distributions at the whole \AGS~ to \LHC~ energy regions}
We use the form of the ~\NUFM~ model as described in Section II and
fit the experimental data with the parameters $y_{e0}$ and $e$ that have
been assumed to be different for different energies, as given in
Table I.  The systematics of these parameters provide useful
information on the collective flow of baryons in these reactions.

Comparing with \NUFM~ calculation before~\cite{feng}, we consider the influence of
baryon number conservation at this time when
discussing the distributions. Fig.1(a) shows net-proton rapidity distributions measured at \AGS~,
and \SPS~ energies. Fig.1(b) shows net-proton rapidity distributions of the top 5$\%$ central collisions measured  at \RHIC~  $\sqrt{s_{\textrm{NN}}}=62.4$
and $\sqrt{s_{\textrm{NN}}}=200$ GeV ,
respectively. The
solid lines are our \NUFM~ calculation results from \AGS~ to \RHIC~
and the dotted line is the calculation result for that of \LHC~. It
can be seen from the Fig.1 that \NUFM~ model can fit  the
experimental results from \AGS~ to \RHIC~, and reproduce central dip of the rapidity distribution of the proton at \SPS~ and
\RHIC~ in agreement with the experimental findings. $y_{e0}$ is
approximately equal to the half width of fit distribution. In the
sense, the parameter $y_{e0}$  represent the kinetic region of
collective flow in the longitudinal direction. The parameter of $T$ is
chosen to be $0.12$ \GeV.

The features of non-uniform flow distributions show strong energy
dependence from \AGS~ to \RHIC~. For example: at \AGS~
($e=0.82,E_{\textrm{lab}}$ =10.8GeV), the net proton distribution has
a peak at mid-rapidity, and the distribution is narrower than that
of the other two energies. The collective flow is approximately
uniform. While at \SPS~~($e=0.61$) a dip begins to show in the
middle of rapidity distribution. While at \RHIC~ $\sqrt{s_{NN}}=62.4$GeV and $\sqrt{s_{NN}}=200$GeV
the distributions show deep dip and the non-uniform parameter $e$ take $0.34$ and $0.31$,respectively.
According to our calculation as the collision energy increases, the net baryon distributions  become wide
for the whole rapidity distribution and the net baryon densities become small at the middle rapidity ($y\approx0$) region.

\begin{figure}[h!]
\centering \resizebox{0.45\textwidth}{!}{
\includegraphics{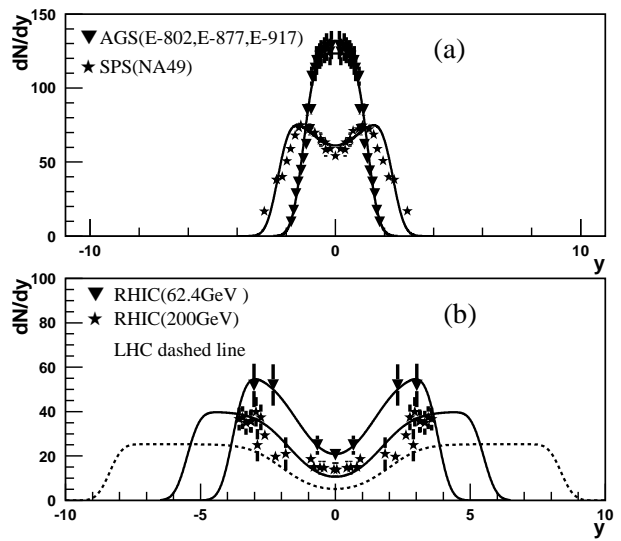}}
\caption{The net proton distribution at \AGS~ and \SPS~ are shown in Fig.1(a).
the net proton distribution at \RHIC~ $\sqrt{s_{NN}}=62.4$GeV and $\sqrt{s_{NN}}=200$GeV are shown
in Fig.1(b). The experimental results come from
~\cite{be1,be2,klay,ahle,ba,ap,videbeck,wienold,back}, the dotted
line which is predicted by \NUFM~ for \LHC~ is shown in Fig.1(b). }
\label{fig1}
\end{figure}

We also speculate the feature of non-uniform flow distributions at ~\LHC~ in fig.1(b).
As mentioned before, although collisions at available heavy-ion energy regions
of \AGS, \SPS~ and \RHIC~ are neither fully stopped nor fully transparent,
but the plateau structure becomes more and more obvious
as the collision energy increases to \SPS~ and \RHIC.
It leads a significant
extension of the kinematic range ($y_{e0}$)in longitudinal rapidity and the net-baryon distribution at the central
rapidity region decreases at ~\LHC. It seems reasonable that we conjecture the kinematic range ($y_{e0}$) at ~\LHC~ $\sqrt{s_{NN}}=5500$ GeV  approaches to the incident beam rapidity $y_{p}$, which is about $1.5$ times of that at ~\RHIC~  $\sqrt{s_{NN}}=200$ GeV ($y_{e0}=5.32$).

By making a analysis of the dependence of $dN/dy\mid_{y=0}$ on incident beam rapidity $y_{p}$ from ~\SPS~($\sqrt{s_{NN}}=17.2$GeV) to ~\RHIC~($\sqrt{s_{NN}}$=62.4 and 200 GeV) experiments, in which the rapidity distribution obviously  show central dip feature, we can provide a relationship between $dN/dy\mid_{y=0}$ and incident beam rapidity $y_{p}$

\begin{equation}  
dN/dy\mid_{y=0}=51.0-22.0\cdot\log(y_{p})  
\end{equation}

\noindent shown in Fig.2.  According to the speculation,  the magnitude of rapidity density at central rapidity $y\approx{0}$ at ~\LHC~ is about $3.63$  which is about  $1/4$ times of that  at ~\RHIC~ 200GeV. We can get  $e=0.19$ to fit the ~\LHC~ distribution
by using the ~\NUFM~ and the baryon number conservation law.

\begin{figure}[h!]
\centering \resizebox{0.45\textwidth}{!}{
\includegraphics{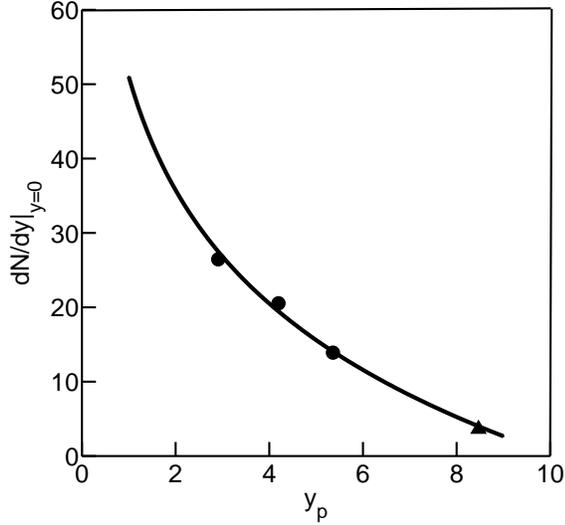}}
\caption{The dependence of central rapidity density $dN/dy\mid_{y=0}$ on incident beam rapidity $y_{p}$ from ~\SPS~($\sqrt{s_{NN}}=17.2$GeV) to ~\RHIC~($\sqrt{s_{NN}}$=62.4 and 200 GeV) experiments. The solid circles are from experimental results, the solid triangle is the speculating .
The real line is the fit curve of Eq.5.
 } \label{fig2}
\end{figure}

At \LHC~, a broad dip in the middle of rapidity region has
developed spanning several units of rapidity, indicating that
collisions are quite transparent at \LHC~ energy region. According
to our study, $e=0.19$   at \LHC~ gives a more obvious non-uniform
feature than that of \AGS~, \SPS~ and \RHIC~ energy region, and the detailed
results are shown in Table.1.

\begin{figure}[h!]
\centering \resizebox{0.45\textwidth}{!}{
\includegraphics{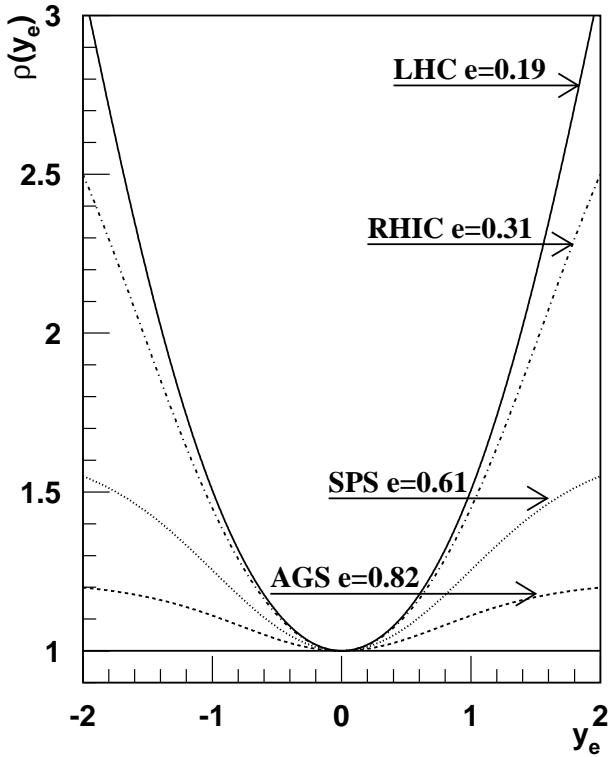}}
\caption{The flow distribution function of net proton in the
longitudinal direction in the whole \AGS~, \SPS~, \RHIC~ and \LHC~
energy regions
 } \label{fig3}
\end{figure}

From Fig.3, we
know that as the incident energy increases, the longitudinal flow
distribution becomes more non-uniform. $e=0.19$ at \LHC~ $\sqrt{s_{\textrm{NN}}}=5500$GeV  is smaller
than $e=0.31$ at \RHIC~$\sqrt{s_{\textrm{NN}}}=200$GeV, and $e=0.82$  at \AGS~ ($E_{\textrm{lab}}$
=10.8GeV). The central rapidity density at ~\AGS~ is the largest in the whole \AGS~, \SPS~, \RHIC~ and \LHC~
energy regions in Fig.3.

Figure 4 shows proton rapidity distribution at different collision energies
at \AGS~ and \SPS, and the solid lines are the calculation results. From Fig.4 we know
that proton distribution shows uniform distribution feature in the longitudinal
direction at the \AGS (2Gev-8GeV) according
to \NUFM. But for Pb + Pb interactions (158GeV) at \SPS, $e=0.61$ shows non-uniform
distribution feature in the longitudinal direction.

Figure 5 shows proton distribution at different collision systems \AGS~ and \SPS. It is found
that from Fig. 5 that $e=1$ and $y_{e0}$=1.411 for heavier collision system (Au + Au), but
$e=0.72$ and $y_{e0}$=1.609 for lighter collision system (Si + Al) at \AGS. It is suggested
that the lighter the collision system, the more non-uniform the distribution in the longitudinal
direction, the larger the kinematical limitation. The same situation is shown at \SPS~ comparing with \AGS.

From the calculation, we find that $y_{e0}$ determines the width
of distribution and confines the flow kinetics regions. $<\beta\gamma>_{\textrm{L}}$

It is found that the depth of the central dip of the net baryon
distributions depends on the magnitude of the parameter $e$ that
describes the non-uniformity of longitudinal flow.

The stopping may be estimated from the rapidity loss experienced by
the baryons in the colliding nuclei. If incoming beam baryons have
rapidity $y_{p}$ relative to the CM, the average rapidity loss of
net proton is

\begin{equation}  
<\delta y>=y_{p}-<y>
\label{eq:eq5} 
\end{equation}

\noindent where $<y>$  is the average rapidity of net proton.

\begin{equation}  
<y>=\frac{2}{N_{\textrm{part}}}\int_{0}^{y_{p}}ydy\frac{dN_{B-\bar{B}}(y)}{dy}
\label{eq:eq6} 
\end{equation}

\noindent where $N_{\textrm{part}}$ is participant nucleon number. $y_{p}$  is
rapidity of incoming beam baryons relative to the CM . The $<y>$
is given by

\begin{equation}  
<y> =
\frac{\int_{0}^{y_{p}}ydy\frac{dn}{dy}}{\int_{0}^{y_{p}}dy\frac{dn}{dy}}
\label{eq:eq7} 
\end{equation}

where $dn/dy$  is given by \NUFM~.

\begin{figure}[h!]
\centering \resizebox{0.45\textwidth}{!}{
\includegraphics{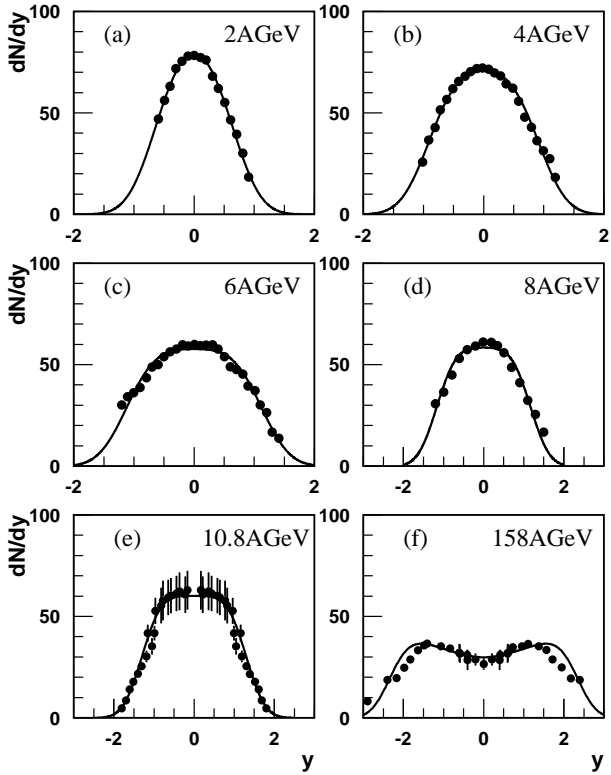}}
\caption{Proton rapidity distributions for Au + Au interaction at AGS.
The experimental data is from Ref.~\cite{ahle,ba,ap,videbeck,wienold,back} and the solid
lines are the calculation results. The whole fitted parameters $e$ and
quanta $y_{e0}$ are given by Table.1}
\label{fig4}
\end{figure}

\begin{figure}[h!]
\centering \resizebox{0.45\textwidth}{!}{
\includegraphics{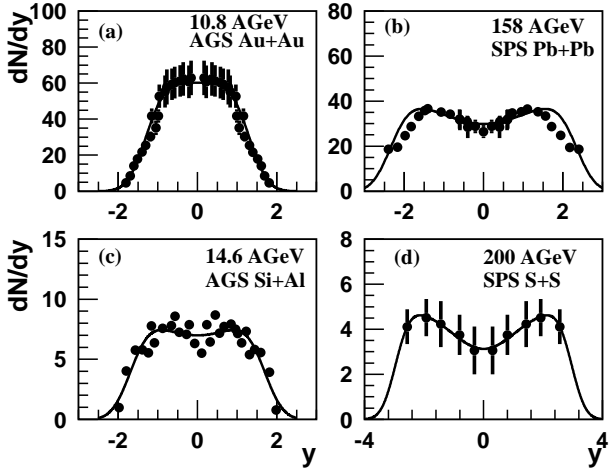}}
\caption{Proton rapidity distributions for Au + Au (10.8GeV) and Si + Al (14.6GeV) interactions at AGS, and for Pb + Pb
(158GeV) and S + S (200GeV) interactions at SPS. The experimental data is from Ref.~\cite{ahle,ba,ap,videbeck,wienold,back} , and the solid lines are the
calculation results. The whole fitted parameters $e$ and
quanta $y_{e0}$  are given by Table.1
 } \label{fig5}
\end{figure}

\begin{table}[h]
\caption{The different parameters of net proton distribution by
using \NUFM~ from \AGS~ to \LHC~}

\centering
\btbl{|c|c|c|c|c|c|}\hline
$E_{lab}$ or $\sqrt{s_{\textrm{NN}}}$ (GeV) & $y_{p}$ & $<\delta_{y}>$ & $<\beta\gamma>_{\textrm{L}}$ & $e$ & $y_{e0}$ \\ \hline
$E_{lab}=2$ (Au+Au AGS) &0.6951&  0.3519&  0.3255&  1.0& 0.648\\ \hline
$E_{lab}=4$ (Au+Au AGS) &1.0647&  0.5391&  0.4653&  1.0& 0.910\\ \hline
$E_{lab}=6$ (Au+Au AGS) &1.2714&  0.6332&  0.5897&  1.0& 1.124 \\ \hline
$E_{lab}=8$ (Au+Au AGS) &1.4166&  0.6997&  0.6189&  1.0 & 1.168 \\ \hline
$E_{lab}=10.8$ (Au+Au AGS) &1.5674&  0.9499&  0.6967&  0.82 & 1.300 \\ \hline
$E_{lab}=14.6$ (Si+Al AGS)  &1.7186&  0.7989&  0.7256&  0.72 &1.684 \\ \hline
$E_{lab}=158$ (Pb+Pb SPS) &2.9112  &1.6774  &1.4558  &0.61 & 2.340 \\ \hline
$E_{lab}=200$ (S+S SPS) &3.0283  &1.1336  &1.6542  &0.554 & 2.960 \\ \hline
$\sqrt{s_{\textrm{NN}}}=62.4$ (Au+Au RHIC)&  4.197& 1.9528&  3.2682  & 0.34 & 4.860 \\ \hline
$\sqrt{s_{\textrm{NN}}}=200$ (Au+Au RHIC)&  5.36& 2.3021&  7.7894  & 0.31& 5.320 \\ \hline
$\sqrt{s_{\textrm{NN}}}=5500$ (Pb+Pb LHC)&  8.4669& 3.5724& 32.7912& 0.19 & 7.880 \\ \hline
 \etbl
\end{table}

\begin{figure}[h!]
\centering \resizebox{0.45\textwidth}{!}{
\includegraphics{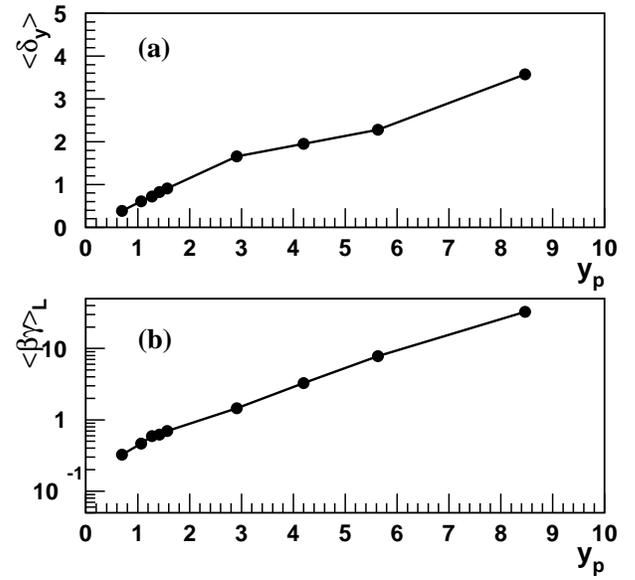}}
\caption{The dependence of average rapidity loss $<\delta y>$  (a)
and $<\beta\gamma>_{\textrm{L}}$ (b) on incident proton rapidity in
the whole \AGS~, \SPS~, \RHIC~ and \LHC~ energy regions.
 } \label{fig6}
\end{figure}

From Fig.6(a) and Table.1, we know that from \AGS~ to \SPS, average
rapidity loss $<\delta y>$  increases linearly with $y_{p}$.
When discussing at \RHIC~, we study  the average rapidity loss at $\sqrt{s_{\textrm{NN}}}=62.4$ and $200$ GeV,  a new linear
increasing relationship is established from \SPS~ to \RHIC~, but
begins to increase slowly and deviates from  that of from \AGS~ to \SPS. We also predict the nuclear stopping power at ~\LHC.
The dependence of $<\beta\gamma>_{\textrm{L}}$ (b) on incident proton rapidity in
the whole \AGS~, \SPS~, \RHIC~ and \LHC~ energy regions are shown in Fig. 6(b). we can find a kind of Log increasing dependence of
$<\beta\gamma>_{\textrm{L}}$ (b) on incident proton rapidity.

\begin{figure}[h!]
\centering \resizebox{0.45\textwidth}{!}{
\includegraphics{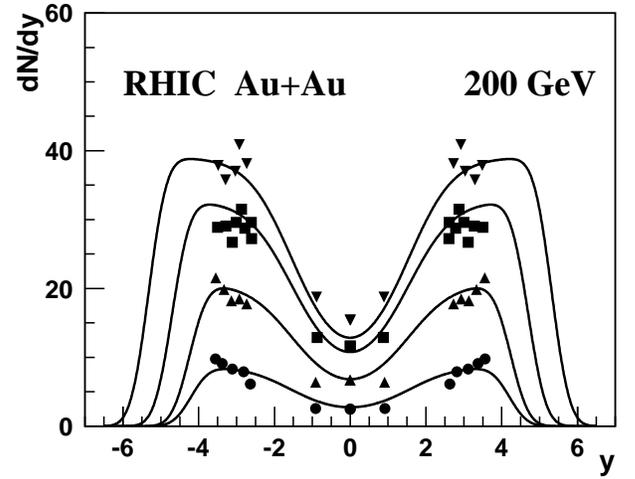}}
\caption{Rapidity distribution of net baryons in Au + Au
collisions at RHIC energy of $\sqrt{s_{NN}}=200$GeV are compared
with preliminary BRAHMS net
baryon data~\cite{Debbe} for different centralities of $0\%$-$10\%$, $10\%$-$20\%$,
$20\%$-$40\%$, and $40\%$-$60\%$. } \label{fig7}
\end{figure}

\begin{table}[h]
\caption{The fit parameters of net proton distribution for
different centralities of  $0\%$-$10\%$, $10\%$-$20\%$,
$20\%$-$40\%$, and $40\%$-$60\%$ by
using \NUFM~ at \RHIC~$\sqrt{s_{\textbf{NN}}}=200$ GeV.}

\centering
\btbl{|c|c|c|c|c|}\hline
centrality at $\sqrt{s_{\textbf{NN}}}=200$ GeV & $y_{e0}$ & $<\delta y>$ & $<\beta\gamma>_{\textrm{L}}$ & $e$ \\ \hline
$0\%$-$10\%$ (Au+Au RHIC) &5.320&  2.318&  7.113&  0.31\\ \hline
$10\%$-$20\%$ (Au+Au RHIC) &4.699&  2.575&  5.195&  0.31\\ \hline
$20\%$-$40\%$ (Au+Au RHIC) &4.239&  2.822&  4.106&  0.31 \\ \hline
$40\%$-$60\%$(Au+Au RHIC) &4.199&  2.844&  4.022&  0.31 \\ \hline
 \etbl
\end{table}

As shown in Figure 7, \NUFM~ can fit the net baryon distribution at different centralities of  $0\%$-$10\%$, $10\%$-$20\%$,
$20\%$-$40\%$, and $40\%$-$60\%$ at  \RHIC~ $\sqrt{s_{\textbf{NN}}}=200$ GeV.
From Figure 8 and Table.2, we know that as the centrality increases, the kinematic region and
average velocity $<\beta\gamma_{\textrm{L}}>$ in the longitudinal direction increases and the distribution becomes wide.
On the other hands, the stopping power diminishes as the centrality increases. It is surprise to find that the non-uniformity
$e$ keeps unchanged.

\begin{figure}[h!]
\centering \resizebox{0.45\textwidth}{!}{
\includegraphics{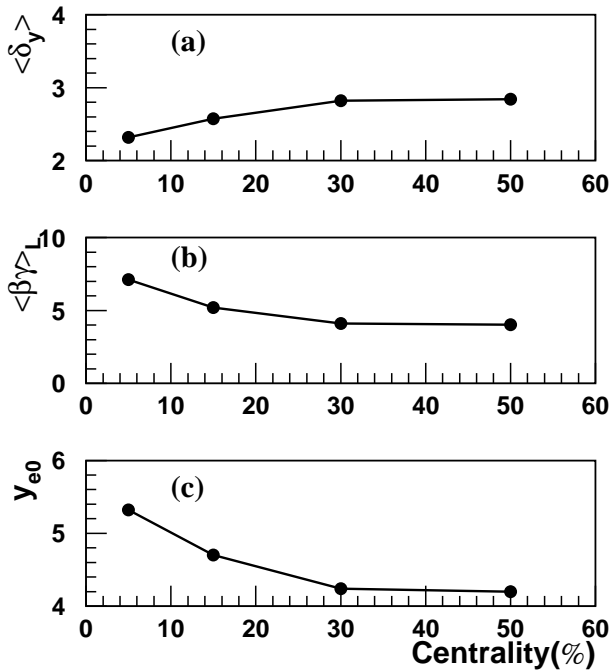}}
\caption{The dependence of average rapidity loss $<\delta y>$  (Fig.8a),
$<\beta\gamma>_{\textrm{L}}$ (Fig8.b) and kinematic region $y_{e0}$ (Fig8.c) on the collision centrality at
RHIC $\sqrt{s_{\textbf{NN}}}=200$ GeV. } \label{fig8}
\end{figure}

\section{Summary and conclusion}

Net proton rapidity distributions have been measured by several experiments at
different energies from \AGS~ to \RHIC~. The compiled data are shown in Fig. 1.
The net proton rapidity distributions are reconstructed among the \AGS~,
\SPS~,\RHIC~ and \LHC~ energy regions by using \NUFM~
in this work. We can predict the distribution feature in the fragmentation
region of the net proton distributions at \RHIC~ although \RHIC~\cite{be1} only
provided the multiplicity distribution of net protons at the central rapidity region.
While at \RHIC~ $\sqrt{s_{NN}}=62.4$GeV and $\sqrt{s_{NN}}=200$GeV
the distributions show deep dip and the non-uniform parameter $e$ take $0.34$ and $0.31$, respectively.
According to our calculation as the collision energy increases, the net baryon distributions  become wide
for the whole rapidity distribution and the net baryon densities become small at the middle rapidity ($y\approx0$) region.

The features of non-uniform flow distributions show strong energy
dependence from \AGS~ to \RHIC~. For example:
the net proton distribution at \AGS~ has
a peak at mid-rapidity, and the distribution is narrower than that
of the other two energies. The collective flow is approximately
uniform. While at \SPS~ a dip begins to show in the
middle of rapidity distribution. The distributions at \RHIC~ $\sqrt{s_{NN}}=62.4$GeV and $\sqrt{s_{NN}}=200$GeV
show the non-uniform feature of deep dip. It is found that the distributions become wider and wider
for the whole rapidity distribution and the densities of net baryon in the middle of rapidity region ($y\approx0$)
become smaller and smaller.

Here,we should mention that quite a few theoretical
models~\cite{Csernai,WongPRL,WongPRD,Hwa,Jeon}
can give equally good representation of the data of particle
productions in relativistic heavy ion collisions. These models
give some different physical pictures for the research. In Ref.~\cite{WongPRD} in order to determine whether
a pure quark-gluon plasma with no net baryon density could be formed in the central rapidity region in relativistic heavy ion collisions,
Wong~\cite{WongPRD} estimated the baryon distribution by using a Glauber-type multiple collisions in which the nucleons of
one nucleus degrade in energy as they make collisions with nucleons in the other nucleus. It was found that in the
head-on collision of two heavy nuclei($A\geq100$), the baryon rapidity rapidity distributions have broad peaks and extend well.

The plateau structure becomes more and more obvious
as the collision energy increases to \RHIC~ although collisions at available heavy-ion energy regions
of \AGS, \SPS~ and \RHIC~ are neither fully stopped nor fully transparent.
It leads a significant
extension of the kinematic range ($y_{e0}$)in longitudinal rapidity and the net-baryon distribution at the central
rapidity region decreases at ~\LHC. We can study the feature of net baryon distributions at \LHC~
by using the ~\NUFM~ and the baryon number conservation law.

Detailed energy dependence of the
net baryon distribution among \AGS~, \SPS~ and \RHIC~ shows a clear transition from
the baryon stopping region to the baryon transparent region. It is found that from \AGS~ to \SPS~,
average rapidity loss $<\delta y>$ increases linearly with $y_{p}$, but begins to increase
slowly and deviates from the linear relationship when at \RHIC~ and
\LHC~.  It may suggests the difference of the interaction mechanism
in \RHIC~ and \LHC~ from \AGS~ and \SPS~. The detailed study
of net proton rapidity distributions from \AGS~ to \LHC~ will deepen our study of
the relativistic heavy-ion collisions.

It is found that the transparency of relativistic heavy-ion
collisions increases as collision energy increases, i.e. the higher
the collision energy, the more transparent of the collision system
by analyzing the proton rapidity distribution. The phase space of
heavy collision system is nearly completely uniform in the
longitudinal direction  at \AGS~. The phase space of proton
distributes non-uniformly in the longitudinal direction at \SPS~ and
\RHIC~. At \LHC~, a broad dip in the middle of rapidity region has
developed spanning several units of rapidity, indicating that
collisions are quite transparent at \LHC~ energy region. According
to our study,  $e=0.19$ at \LHC~ gives a more obvious non-uniform
feature than that of \AGS~, \SPS~ and \RHIC~ energy region. By
reanalyzing \RHIC~, we obtain a wider rapidity distribution than
that of Ref~\cite{feng}.

\section{Acknowledgments}
This work was supported by National Natural Science Foundation of China (10975091),
Excellent Youth Foundation of Hubei Scientific Committee (2006ABB036)and Education Commission of Hubei
Province of China (Z20081302).The authors is indebted to
Prof. Lianshou Liu for his valuable discussions and very helpful
suggestions.

{}

\end{document}